\documentclass[dvips,10pt,prl,aps,nofootinbib,twocolumn]{revtex4-1}
\usepackage{amsmath}
\usepackage[hmargin=2.15cm,vmargin=2.15cm]{geometry}

\newcommand\myeq{\mathrel{\overset{\makebox[0pt]{\mbox{\normalfont\tiny\sffamily def}}}{=}}}

\begin{document}

\author{Steven Willison}

\address{Centro Multidisciplinar de Astrofisica - CENTRA, Departamento de F\'{i}sica,
Instituto Superior T\'{e}cnico - IST, Universidade de Lisboa - UL, Av. Rovisco Pais 1, 1049-001 Lisboa, Portugal}

\email{steven.willison@ist.utl.pt}

\title{Local well-posedness in Lovelock gravity}
\date{09 January 2015}

\begin{abstract}
It has long been known that Lovelock gravity, being of Cauchy-Kowalevskaya type, admits a well defined initial value problem for analytic data. However, this does not address the physically important issues of continuous dependence of the solution on the data and the domain of dependence property. In this note we fill this gap in our understanding of the (local) dynamics of the theory. We show that, by a known mathematical trick, the fully nonlinear harmonic-gauge-reduced Lovelock field equations can be made equivalent to a quasilinear PDE system. Due to this equivalence, an analysis of the principal symbol, as has appeared in recent works by other authors, is sufficient to decide the issue of local well-posedness of perturbations about a given background.
\end{abstract}

\maketitle

Lovelock gravity is a candidate modified gravity in higher dimensions\cite{PK}. A key feature is that it is the most general metric theory with second derivative field equations consistent with covariant conservation and symmetry of the stress tensor. When motivating the theory from a theoretical standpoint, one might say that the Cauchy problem is ``the same as in GR" in some sense. To make precise in what sense this is true is an important but somewhat neglected question.

The matter was studied in Refs. \cite{Aragone,CB} and related issues in the Hamiltonian formulation in Ref. \cite{TZ} (for more recent developments on the Hamiltonian formalism see e.g. Refs. \cite{Kunstatter:2012jr}). There was a long hiatus until very recently, when the subject was revived \cite{Izumi,Reall1,Reall2} and the characteristics of the theory were looked at carefully for physically interesting backgrounds. For example, it was found that Killing horizons were always characteristic surfaces in Einstein-Gauss-Bonnet\cite{Izumi} and also in higher Lovelock theory\cite{Reall1}.

There are various alternative definitions of hyperbolic for nonlinear systems. The definition in the above mentioned articles is that the system is hyperbolic with respect to a given background solution if its linearisation is hyperbolic. Sometimes in PDE theory the local well posedness of the system, for an appropriate function space, is taken to be the defining property\footnote{For our purposes (local) well-posedness means there is a region $M \cong \Sigma \times [0,T]$ such that for any initial data in the appropriate regularity class on $\Sigma$ there is one and only one solution on $M$ and that
solution depends continuously on the data.}. However, if the nonlinear terms contain products of second derivatives, the latter property does not automatically follow from the former definition\footnote{This problem arises in a method of successive approximations to prove existence and continuous dependence of solutions for non-analytic data. If the principal coefficients of the linearised equation for the $n$-th order approximation are second derivative in the $(n-1)$-th order solution, there is a loss of differentiability at each step, rendering the Banach fixed point theorem inaplicable. Therefore, standard well-posedness results apply only to quasilinear systems.}.
Therefore the authors of \cite{Reall1} stopped short of making positive claims about well-posedness.

In this note, we show that it is possible to make such affirmations. By a known mathematical trick we can convert the gauge reduced Lovelock equations into a quasilinear PDE system. This trick works essentially for the same reason that the equations can be brought to Cauchy-Kowalevskaya form- there are no squares of second time derivatives. By taking spatial derivatives and defining new variables accordingly one obtains the required quasilinear system. This trick is well-known in other contexts e.g. Ref \cite{HanHong} pg. 62, but seems to have been overlooked as regards modified gravity theories.


In the case of General relativity, in order to have well-posedness it is necessary to gauge-fix the diffeomorphisms. Also one must restrict the initial data to obey constraint equations. Then one obtains a well-posed problem. The gauge and constraint structure of Lovelock theory is essentially the same, and the same strategy can be applied.
The reduced Lovelock equations w.r.t. harmonic gauge are, in a local coordinate chart \cite{CB}:
\begin{gather}
 g^{\rho\sigma} \partial_\rho\partial_\sigma g_{\mu\nu} = \lambda Z_{\mu\nu}(g,\partial g, \partial \partial g) + \cdots
\end{gather}
Here $\lambda$ is some coupling coefficient multiplying the higher ($n>1$) order Lovelock terms. Here and in what follows $\cdots$ denotes terms of order less than 2 in derivatives of the basic fields. The only relevant property of $Z$ is linearity in second time derivatives.

Let us now privilege the second time derivatives
\begin{gather}
 g^{00} \partial_0\partial_0 g_{\alpha\beta} \left( \delta_{(\mu\nu)}^{(\alpha\beta)} + \lambda X_{\mu\nu}^{\alpha\beta} \right)
 = L_{\mu\nu} + \lambda Q_{\mu\nu}
\end{gather}
where $X$, $L$, $Q$ depend on second derivatives but not on $\partial_0\partial_0 g$. $L_{\mu\nu} = -2g^{i0} g_{\mu\nu, i 0} -g^{ij}g_{\mu\nu,ij} +$ lower derivative terms.
Let us assume that $\delta +\lambda X$, viewed as a matrix of rank $n(n+1)/2$, is invertible and write this inverse as
$(\delta +\lambda X)^{-1} \myeq \delta +\lambda Y$.
Then
\begin{gather}
 g^{\rho\sigma} \partial_\rho\partial_\sigma g_{\mu\nu} = \lambda Y_{\mu\nu}^{\rho\sigma} L_{\rho\sigma} + \left(\delta_{(\mu\nu)}^{(\rho\sigma)} +\lambda Y_{\mu\nu}^{\rho\sigma}\right)\lambda Q_{\rho\sigma} + \cdots \label{start}
\end{gather}
Now we may consider the system of (\ref{start}) along with its first spatial derivatives, introducing an auxiliary field through the replacement  $\partial_i g_{\mu\nu} \to v_{\mu\nu i}$. We have
\begin{gather}
 g^{\rho\sigma} \partial_\rho\partial_\sigma g_{\mu\nu} = \cdots \label{system1}
\\
 g^{\rho\sigma} \partial_\rho\partial_\sigma v_{\mu\nu i} -\lambda A^{jk\mu\nu}_{\alpha\beta} \partial_k \partial_i  v_{\mu\nu j}
 -\lambda B^{j \mu\nu}_{\alpha\beta} \partial_j \partial_0  v_{\mu\nu i} = \cdots\label{system2}
\end{gather}
(where
\begin{gather}
 A^{jk\mu\nu}_{\alpha\beta}  := \frac{\partial}{\partial  v_{\mu\nu j,k} } ( Y L + (\delta + \lambda Y) Q)_{\alpha\beta}
 \\
 B^{j \mu\nu}_{\alpha\beta} := \frac{\partial}{\partial  v_{\mu\nu j,0} } (Y L + (\delta + \lambda Y)  Q)_{\alpha\beta}
\end{gather}
are first derivative in the fields)
together with initial value constraints  $\phi_{\mu\nu i} := \partial_i g_{\mu\nu} - v_{\mu\nu i} = 0$,  $\partial_0\phi_{\mu\nu i}= 0$ which are preserved by the evolution.
(\ref{system1}) and (\ref{system2}) constitute a quasilinear PDE system. It is equivalent to the reduced Lovelock equations provided that $\delta +\lambda X$ is invertible and that $g^{00} \neq 0$ (i.e. a $t=$ constant surface is not characteristic and non-null respectively.).

Verifying hyperbolicity about a given background would therefore guarantee local well-posedness for an appropriate function space, by well-known results. 
The system is not quasidiagonal, so some work is required. But, since $A$ and $B$ are order $\lambda$, for small perturbations about flat spacetime, the system is obviously hyperbolic and the characteristics are approximately light-like as expected. Since our system is only equivalent to the covariant Lovelock equations modulo constraints, we expect to have some spurious characteristic cones for pure gauge degrees of freedom\cite{CB}. For general backgrounds,  a study of the original equations using the gauge invariant approach advocated in Ref. \cite{Reall1} will probably be more useful and illuminating than a direct study of (\ref{system1}-\ref{system2}).

Regarding the choice of function space, we note that in this construction we introduce an auxiliary variable which is the spatial derivative of the metric. So we expect existence and uniqueness proofs to require 1 extra order of differentiability of the initial data compared to Einstein theory, which could spell trouble for low regularity solutions like thin shells.
For example, we can not rule out the possibility of the shell evolution being non-unique due to the kind of branching solutions discussed in Ref. \cite{Ramirez}.
Therefore, well-posedness of the braneworld model for Lovelock theory remains in doubt.

Questions related to the global Cauchy problem have not been treated here but it is important to mention the discussion in Ref. \cite{Reall2}. Based on a study of shock formation, it was argued that weak cosmic censorship could be violated. It was suggested that Minkowski space should be stable, based on analogy with GR and the wave equation, and on general principle that in higher dimensions there is more dissipation. To put this on a firmer footing one would need to show that qualitatively different terms appearing in the higher order corrections do not lead to amplification. The quasilinear reformulation presented here may perhaps be helpful in that context, but obtaining exact results seems to be a formidable problem.

\acknowledgements We thank Jorge Silva for a helpful discussion. This work was supported by the
Funda\c{c}\~{a}o para a Ci\^{e}ncia e a Tecnologia of Portugal and
the Marie Curie Action COFUND of the European Union Seventh
Framework Programme (grant agreement PCOFUND-GA-2009-246542).

\end{document}